\documentstyle[pra,aps]{revtex}
\draft
\title{Realization scheme for continuous fuzzy measurement of energy\\ 
and the monitoring of a quantum transition\thanks{quant-ph/9808062}}

\author{J\"urgen Audretsch\cite{bylineaud}
}
\address{Fakult\"at f\"ur Physik der Universit\"at Konstanz,\\
Postfach 5560 M 674, D-78434 Konstanz, Germany}

\author{Michael Mensky\cite{bylinemens}
}
\address{P. N.Lebedev Physical Institute, 117924 Moscow, Russia\\
and\\
Fakult\"at f\"ur Physik der Universit\"at Konstanz,\\
Postfach 5560 M 674, D-78434 Konstanz, Germany}

\date{December 4, 1998}

\begin{document}
\maketitle

\begin{abstract}
The continuous fuzzy measurement of energy of a single two-level 
system driven by a resonant external field is studied. An analysis is 
given in the framework of the phenomenological restricted path 
integral approach (RPI) (which reduces effectively to a Schr\"odinger 
equation with a complex Hamiltonian) as well as with reference to the 
microphysical details of a class of concrete physical realizations. 
Within the RPI approach it is demonstrated that for appropriately 
adjusted fuzziness, information about the evolution of the state of 
the system $|\psi(t)\rangle$ can be read off from the measurement 
readout $E(t)$. It is shown furthermore how a measurement of this type 
may be realized by a series of weak and short interactions of the 
two-level system with a quantum mechanical meter system. After each 
interaction a macroscopic measuring apparatus causes the meter to 
transit into one of two states. The result is used to generate the 
energy readout $E(t)$. In this way a complete agreement with the RPI 
approach is demonstrated which thus obtains an operational 
interpretation.
\end{abstract}



\section{Introduction}

The measurement of a physical quantity $A$ on a quantum system, as it
is introduced as a well known postulate of quantum mechanics, is
based on a {\it single, instantaneous} measurement which is sharp and
has a strong back influence on the system. Restricting for simplicity
to a non-degenerate discrete spectrum, to a pure state and to a
selective measurement, then {\it sharp} (or {\it perfect}) means that
the measurement yields exactly an eigenvalue $a_n$ of the observable
$A$ as the only possible result. That the measurement is {\it hard}
(or its influence is {\it strong}) means that it results in a
transition of the state of the system to the respective eigenvector of
the observable {\it (projection measurement)}. For the probability of
finding the result $a_n$, reference is made to a large number of
single measurements on equally prepared states {\it(ensemble)}.

In contrast to this we want to discuss in the following
a general class of quantum measurements. 
Instead of a single instantaneous measurement,
we consider a {\it continuous} measurement
performed on one and the same {\it single} quantum
system over a certain time leading to a measurement readout
which is a function of time $a(t)$.
The influence on the system is
allowed to be {\it weak}
(or {\it the measurement to be soft}) which means that the
system is not transferred to an eigenstate and that the readout must
not agree with an eigenvalue. Accordingly a weak measurement does not
discriminate between eigenvalues. In this sense the measurement is
{\it unsharp} (or of {\it low resolution}).
The readout $a(t)$ need not to be close to the eigenvalues $a_n$.
During short intervals the readout curves may even show large
deflections from $a_n$. In addition for
the same initial state and the same continuous measurement, there may
result different readouts, with a certain probability
distribution. Because of
these characteristic properties of this type of measurement, which
originate from its
quantum nature, we call the whole class continuous {\it fuzzy}
measurements.

In the last decades continuous quantum measurements were considered
intensively
\cite{ContinMeas,M79,book93,KhalZeno88,BruHar90,Schenzle92,%
BragKhalBk,Zeno,Milburn,%
GisinKnightPercivalThompson,JacobsKnight,Peres-book,Onofrio-energy}.
We mention that such a measurement is closely related with the
phenomenon of decoherence.
In the context of the emergence of classical properties
this phenomenon has recently been discussed in \cite{ZehBk} where also
an extensive survey of the literature may be found.

The realization of a continuous measurement
may consist of a series of short measurements during
which the quantum system interacts with the measurement device.
If this interaction is weak in a specific sense the state of the system does
not transit to an eigenvector
of the respective observable.
In the intervals between, the system follows its internal dynamics.
Depending on the weakness of the measurement interaction
and the lengths of the
intervals between, fuzziness may be obtained.
In this paper we describe the underlying physical
microstructures in detail.

A particular limiting case of our general class of continuous measurements
is the well known quantum {\it Zeno effect} which has been
predicted and then confirmed experimentally \cite{Zeno}.
The content of this effect is that a continuous measurement which is sharp
prevents transitions between discrete spectrum states (for
example atom levels).
The reason for this strong influence is that the considered
continuous measurement is in fact approximated by a series of very quickly
repeated short measurements which are so strong that each of them
projects the state of the system on one and the same discrete eigenstate.
In \cite{JacobsKnight} such a procedure is called a ``continuous
projection measurement". 
The readout agrees constantly with the respective eigenvalue.
In practice the influence on the measured system does
not lead to an absolute freezing of its evolution
but still modifies it considerably \cite{Milburn}. In the specific
setup of a three-level system a random telegraph-type signal may be
obtained from the measured system as a consequence of its "shelving"
\cite{GisinKnightPercivalThompson}. The Zeno effect thus demonstrates 
in a most evident way
that the back reaction during a continuous measurement may strongly
influence the measured system.

This motivates us to answer the following complementary questions: Is 
it possible to measure continuously a single quantum system in such a 
weak way that the behavior of its state does not radically differ from 
what it would be if no measurement is performed (this could be for 
example a Rabi oscillation)? And is in the case of such a fuzzy 
measurement to be expected that the obtained continuous measurement 
readout reflects the motion of the state of the system thus making it 
visible? To answer these questions is the first aim of this paper.

In the context of an ensemble approach (non-selective description), 
the continuous fuzzy measurement of an observable
(energy) with discrete spectrum has been investigated in
\cite{KhalZeno88,BragKhalBk,Peres-book}. In
\cite{KhalZeno88,BragKhalBk} the probability for transition
from one level to another was determined as an average over the
ensemble of many identical measured systems (equivalently, over
the set of all possible measurement readouts). In \cite{Peres-book}
the measurement readout averaged over the ensemble (presented by 
a density matrix) was shown to correlate with the Rabi oscillations 
between the energy levels.

However this approach does not give a complete picture of the
behavior of an individual system like a single atom which
is continuously measured.
Particularly, our question what are the dynamics of the system
subject to a measurement with a certain readout cannot be
answered in the framework of the ensemble approach .
This requires the selective description of the measurement.

The phenomenological restricted-path-integral (RPI) approach \cite{M79,book93}
to continuous fuzzy measurements sketched in Section~\ref{SectPhenomen}
provides such a selective description. It takes into account
not only the back reaction onto the system caused by the measurement,
but also specifies how it is determined by the information
obtained in the concrete measurement readout.
For a review of different phenomenological approaches see \cite{POTrev96}.
The RPI approach, which is equivalent to the use of a
complex Hamiltonian, has first been used for continuous measurements of
observables having continuous spectra. It also has been applied to a
discrete-spectrum observable (energy of a two-level system) in
\cite{Onofrio-energy}.
There it was shown that the quantum Zeno effect arises if the resolution
of the measurement is good enough (in comparison with the level
difference).
However the analysis given in \cite{Onofrio-energy} was not complete
because only the special case of constant measurement readouts
coinciding with the energy levels $(E(t) = E_n = {\mathrm{const}})$ was
considered.
A detailed analysis of the measurement of energy of a two-level
system with arbitrary measurement readouts $E(t)$ taken into account,
so that fuzziness is incorporated, has been performed in \cite{AudM97}.

This analysis confirmed that the behavior of the system is of Zeno-type
with level transitions frozen if the resolution of the measurement
is high (hard measurement).
The opposite regime of poor resolution or very fuzzy (soft)
measurements has also been investigated.
In this case the measurement influences the internal dynamics
of the measured system only insignificantly.
As a consequence of the low resolution the probability distribution of all
possible readouts is very broad and a single readout does not contain
reliable information about the systems evolution.

The most interesting regime of measurement is the intermediate
regime. Preliminary results obtained in \cite{AudM97,AuMNam97}
show that in this case,
when the resolution is not too low to give no information
and not too high to lead to the Zeno effect, induced transitions
between levels (for example Rabi oscillations) maintain though
they are modified, and the measurement readout $E(t)$ is correlated
with these oscillations thus making them visible.
Below we will improve these studies in Section~\ref{monitoring}.

The phenomenological approach, although it seems to be
convincing in itself, demands a microphysical justification.
It is therefore the second aim of this paper to specify the details of
the dynamics caused by a class of interactions of the two-level
system with a subsidiary system which reproduce exactly the
results predicted by the RPI approach to a continuous fuzzy
measurement of energy. We then know how such a measurement of energy
can be realized operationally and for which situations the
phenomenological scheme may be used in practice. In
Section~\ref{realization} we study the underlying dynamics
and discuss a particular type of interaction. This generalizes an
example given in \cite{AuMNam97}. As far as experimental realizations 
are concerned, it is also a generalization of the 
quantum optical setup proposed in \cite{BruHar90} and 
considered in more detail in \cite{Schenzle92}. But the aim of the latter 
two papers was not a continuous measurement in it own right but 
the decoherence of levels resulting from the long enough continuous 
measurement. 

\section{Phenomenological approach to continuous measurement}
\label{SectPhenomen}

In a first step we summarize very briefly the phenomenological and therefore
model-independent RPI approach to continuous fuzzy measurements.
The aim of this scheme is to include the effect
of measurement on the dynamics of a {\it single} quantum system.
The measuring apparatus and its interaction with the system
is thereby only implicitly taken into account.
It turns out that it is enough to know what information about the system
the measurement supplies.
Below we formulate the theory directly with respect to the
measurement of the observable
energy. The more general scheme and additional details
are described in \cite{AudM97}.

We discuss a system with a Hamiltonian $H=H_0+V$ where $H_0$ is the
Hamiltonian of a "free" multilevel system and $V$ is an external
potential. Below we will assume that it acts as a driving field
leading to transitions between levels. The continuous fuzzy
measurement of energy during the time interval $[0,T]$ results in a
curve $[E]=\{E(t) | 0 \le t \le T\}$ called the {\it measurement
readout}.

The curve $[E]$ presents the information supplied by
one specific continuous measurement process.
Because $[E]$ is assumed to be known we deal with a selective
description of the measurement.
Our scheme has then to answer the following questions: Given an
initial state $|\psi_0 \rangle$ of the system for $t=0$ and a particular
readout $[E]$, what is i) the probability density $P[E]$ to obtain this
readout in a measurement and ii) what will be the state
$|\psi_T^{[E]} \rangle$
into which the system is transferred at the end of the
measurement at time $t=T$.
The respective predictions can be made in the formalism
of restricted path integrals (RPI) which leads to dynamics with
complex Hamiltonian.

\subsection{Restricted path integrals}

We turn to the path integral formulation of continuous
measurements.
The information supplied by the measurement with readout $[E]$ is
represented by a weight functional $w_{[E]}[p,q]$ which serves
to restrict (or weight) the different paths contributing to the
propagator according to

\begin{equation}
U_T^{[E]}\, (q'',q')
=\int d[p]d[q]\,w_{[E]}[p,q]\,
e^{\frac{i}{\hbar}\int_{0}^{T} (p\dot q - H(p,q,t))}.
\label{part-prop}\end{equation}
The concrete form of the weight functional may be different depending
on the type of the measurement. The simplest choice is a
{\it Gaussian functional}

\begin{equation}
w_{[E]}[p,q] = \exp\left[ -\kappa \int_{0}^{T}
\Big( H_0\big(p(t),q(t),t\big) - E(t) \Big)^2\,dt \right]
\label{weight}\end{equation}
which becomes exponentially small if the square average deflection

\begin{equation}
\langle (H_0-E)^2\rangle_{T}
= \frac{1}{{T}}\int_{0}^{T}\Big( H_0\big(p(t),q(t),t\big) 
- E(t) \Big)^2\,dt
\label{deviation}\end{equation}
between the curves $[H_0]$ and $[E]$ is large. The parameter $\kappa$
represents the strength of the influence of the continuous measurement
on the system.
Its inverse may serve as a {\it measure of fuzziness}. It is one of 
the aims of this paper to describe general properties of measurement 
setups and their realization for which choice (\ref{weight}) is 
appropriate. Eq.~(\ref{weight}) can be written in the form

\begin{equation}\label{weight-energy}
w_{[E]}[p,q]
= \exp\left[
-\frac{\langle (H_0-E)^2\rangle_{T}}{{{\Delta E}}^2_{T}}\right]
\end{equation}
where the parameter
\begin{equation}\label{DeltaE}
{{\Delta E}}_{T} = \frac{1}{\sqrt{\kappa {T}}}
\end{equation}
is introduced characterizing the fuzziness of the measurement in a more
transparent way than $\kappa$. It has been shown in \cite{AudM97} that it
represents the measurement resolution obtained after the time $T$.
Its improvement with time reflects the remaining
Zeno-type influence of a fuzzy measurement carried out continuously.

If the continuous fuzzy measurement results in a certain readout $[E]$,
then the state of the system has changed after time $T$ according to:
\begin{equation}
|\psi_T^{ \left[ E\right]}\rangle=U_T^{ \left[ E\right]}|\psi_0\rangle,
\quad\varrho_T^{[E]}=U_T^{[E]}\varrho_0 \left(U_T^{[E]}
\right)^{\dagger}.
\label{evolut}\end{equation}

These formulas contain 
partial evolution operators depending on $[E]$.
The first of these formulas may be applied to the evolution of pure
states to which we will restrict below while the second is valid 
both for pure and mixed states.
Note that because of the inclusion of the weight factor the norm of
the resulting state vector $|\psi_T^{[E]} \rangle$ is less than unity.

Normalization may of course be performed, but it is much more convenient
and transparent to work --- as we will do --- with
{\it non-normalized states.}
Their norms give directly according to \cite{book93} the
{\it probability densities}

\begin{equation}
P \left[ E\right]=\langle\psi_T^{[E]}|\psi_T^{[E]}\rangle
\label{prob}\end{equation}
of the corresponding measurement readout $[E]$. The time dependence of the 
state vector $|\psi_t\rangle $ shows how the system evolves in case 
the measurement gives the readout $[E]$. Equations (\ref{prob}) and 
(\ref{evolut}) answer therefore the questions i) and ii) asked 
above.\footnote{A finite probability never corresponds to a single readout 
$[E]$ but to a family of readouts. It may be found by integration of the probability 
density (\ref{prob}) over this family \cite{AudM97}. The corresponding behavior 
of the system is given by the integral (over the same family) of the second of 
Eqs.~(\ref{evolut}). From physical point of view this means that it is impossible 
to know $E(t)$ at each instant $t$ precisely, but only with some finite accuracy. 
Description of the measurement in this more realistic situation is partly 
non-selective and leads to a mixed state of the measured system. However, 
the description of the system behavior by time-dependent pure state 
$|\psi_t^{[E]}\rangle$ (selective description) gives a good approximation 
if the measurement readout is well controlled (the above-mentioned family 
of curves $[E]$ is a thin bundle). It becomes precise if we deal not with 
actually continuous measurement but with a sequential (repeated) 
measurement of a non-degenerate discrete-spectrum observable. 
This is the case in the model to be considered below, in 
Sect.~\ref{realization}.}

\subsection{Complex Hamiltonian}

The restricted path integral (\ref{part-prop}) with the
weight functional (\ref{weight}) has the form of a
non-restricted Feynman path integral but with an {\it effective
Hamiltonian}
\begin{equation}\label{effect-Ham}
H_{[E]}\,(p,q,t) = H(p,q,t) - i\kappa\hbar \,\big( H_0(p,q,t) - E(t)
\big)^2
\end{equation}
containing an imaginary term. This means that for a particular given 
readout $[E]$ the evolution of an initial state $|\psi_0 \rangle$ may 
as well be obtained by solving the corresponding Schr\"odinger 
equation

\begin{equation}
\frac{\partial}{\partial t} |\psi_t\rangle
  = \left(-\frac{i}{\hbar} H
  -\kappa \,\Big( H_0 - E(t)\Big) ^2\right)\, |\psi_t\rangle.
\label{efSchroedEnergy}\end{equation}
The dynamics accompanying a continuous fuzzy measurement is therefore 
reduced to an equivalent Schr\"odinger equation with a complex 
Hamiltonian. The appearance of an imaginary part indicates as a 
characteristic trait of the process of continuous measurement that 
information about the system is dissipated in its environment.

In the following we will restrict to a {\it 2-level system.}
We assume that it is under resonant influence of a periodical force
with frequency $\omega =\Delta E/\hbar$ (may be with a slowly
altering amplitude), where $\Delta E=E_2-E_1$.
Expanding the state $|\psi_t\rangle$ in the basis

\begin{equation}\label{eigenstates}
|\varphi_n(t)\rangle = e^{-iE_n\, t/\hbar}|n\rangle
\end{equation}
of time-dependent eigenstates of $H_0$, we have the following
system of equations for the coefficients of the expansion 
$|\psi_t\rangle = \sum C_n(t)|\varphi_n(t)\rangle$:

\begin{eqnarray}
\dot C_1 &=&  -i v(t) C_2 - \kappa (E_1-E(t))^2\, C_1,
\nonumber\\
\dot C_2 &=&  -i v(t) C_1 - \kappa (E_2-E(t))^2\, C_2
\label{2levelEq}\end{eqnarray}
where $v$ is proportional to the matrix elements

\begin{equation}
\hbar v=\langle\varphi_1|V|\varphi_2\rangle
=\langle\varphi_2|V|\varphi_1\rangle^{*}
\label{driv-field}\end{equation}
which are constant or alter slowly as compared with the frequency 
$\omega$.

\section{Monitoring a level transition by the corresponding energy 
readout} \label{monitoring}

We want to discuss the question whether the time development of the 
state can be made visible by the measurement readout. Let us describe 
first the case when there is no continuous measurement $(\kappa =0)$ 
and the wave function is normalized. Then due to the driving field 
$V$ the state of the system undergoes Rabi oscillations between the 
eigenstates with Rabi period $T_R=\pi/v$. We consider a $\pi$-pulse 
of the driving field in the interval $[T_1, T_2]$ where $0 \le 
T_1 \le T_2 \le T$, so that $T_2-T_1=T_R/2$ and $v$ in 
(\ref{2levelEq}) vanishes outside of this interval. If the system is 
at the time $T_1$ on level 1 $(C_1(T_1)=1,\,\, C_2 (T_1)=0)$, then the 
system is subject to the level transition during the interval $[T_1, 
T_2]$. This means that the probability ${|C_2(t)|}^2$ to be at level 2 
gradually increases during this interval and achieves unity at the end 
of it. The system stays at level 2 also after $t=T_2$ and therefore it 
is with certainty at level 2 at $t=T$. We shall see how this process 
is modified if the continuous measurement of energy is performed.

A quantity by which in this case the fuzziness of this measurement can 
be characterized and which may replace $\kappa$ in physical 
discussions is the {\it level resolution time} obtained from 
(\ref{DeltaE}):

\begin{equation}
T_{\mathrm lr} = \frac{1}{\kappa\Delta E^2}.
\label{Tlr}\end{equation}
If a continuous fuzzy measurement lasts much longer than $T_{lr}$, it 
distinguishes (resolves) between the levels \cite{AudM97}. Small 
$T_{lr}$ represents quick level resolution because of small fuzziness 
and therefore strong Zeno-type influence of the measurement. Such a 
measurement may be called {\it hard} and the corresponding regime {\it 
Zeno regime}. The larger $T_{lr}$ is, the weaker is the influence of 
the measurement on the atom. The Rabi oscillations may show up. A 
large $T_{lr}$ corresponds therefore to a {\it soft} measurement ({\em 
Rabi regime}).

It has been shown in \cite{AudM97} that for $T_{lr}$ being of the 
order of $T_R/4 \pi$ there is a regime in which significant 
correlations between oscillations of the state vector and the energy 
readout $[E]$ exist. This intermediate regime lies between the Zeno 
and Rabi regimes. In extending \cite{AudM97} we present now some 
results of a numerical analysis of Eq.~(\ref{2levelEq}) for all three 
regimes of measurement: Zeno-type, Rabi-type and intermediate. A 
sketch of the main results in the intermediate regime has already been 
published in \cite{AuMNam97}.

Assuming the influence of a continuous measurement of energy with given
fuzziness on a quantum system, we want to treat the following problems:

\renewcommand{\labelenumi}{\roman{enumi})}
\begin{enumerate}
\item For a given readout $[E]$, what is the corresponding development 
of the state $|\Psi_t\rangle$ of the system as it is represented by 
the modulus $|c_n(t)|$ of the normalized coefficient $c_n(t) = C_n(t)/ 
\sqrt{|C_1(t)|^2 + |C_2(t)|^2}$? We denote the whole curve ${\cal 
P}_2(t) = |c_n(t)|^2$ by $[{\cal P}_2]$. ${\cal P}_2(t)$ is the 
probability to find the system in the state 2 if a projection 
measurement would be made at the time $t$.

\item In addition we want to know the probability density
$P[E]$ to obtain a particular readout $[E]$ which is simultaneously 
a probability density for the related state curve $[{\cal P}_2]$.
\end{enumerate}

In the course of the numerical analysis random curves $E(t)$ were 
generated, for each of them Eq.~(\ref{2levelEq}) was solved, the norm 
of the solution at the final time $|C_1(T)|^2 + |C_2(T)|^2$ was 
calculated  and according to (\ref{prob}) interpreted as the 
probability density $P[E]$ that the curve $E(t)$ occurs as the 
measurement readout. The related state curve ${\cal P}_2(t)$ was 
worked out. Being determined as soon as $[E]$ is fitted, $[{\cal 
P}_2]$ carries the same probability density as $[E]$. For further 
analysis all curves $E(t)$ were smoothed with the time scale of the 
order of $T_2-T_1$ to eliminate insignificant fast
oscillations but conserve information about the transition. We discuss 
the results for the different regimes separately.

\subsection{The two limiting cases}

\begin{figure}
\let\picnaturalsize=N
\def\picsize{3.0in}
\def\picfilename{gausfig1.eps}
\ifx\nopictures Y\else{\ifx\epsfloaded Y\else\input epsf \fi
\let\epsfloaded=Y
\centerline{\ifx\picnaturalsize N\epsfxsize \picsize\fi \epsfbox{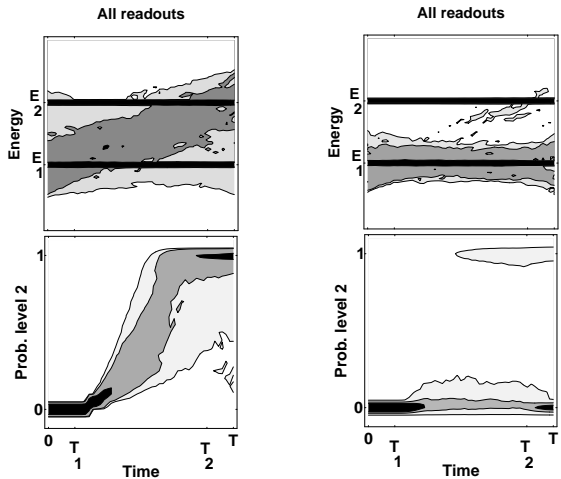}}}\fi
\caption{\label{Fig1}Density plots of measurement readouts $E(t)$
and the corresponding probability curves ${\cal P}_2(t) = |c_2(t)|^2$
characterizing the behavior of the state of the system in time
for two values of fuzziness: a soft measurement with
$4\pi T_{lr}/T_R= 10/3$ (left) and $4\pi T_{lr}/T_R=2/3$ (right).}
\end{figure}

The results of the analysis for the two limiting choices of fuzziness 
are given in the density plots of Fig.~\ref{Fig1}. The 
smoothed curves $E(t)$ are presented in the upper diagrams and the 
corresponding curves ${\cal P}_2 (t)$ in the lower diagrams. The 
degree of grayness in each point of the diagram indicates the 
probability density that a curve passes through this point. The figure 
thus gives an idea of the limits in which a relatively probable 
measurement readout must lie.

The left part of Fig.~\ref{Fig1} refers to a soft continuous 
measurement. It belongs to the Rabi regime where the influence of the 
driving field dominates. Regarding the development of the state, the 
measurement only slightly influences the measured system not 
preventing the level transition which occurs with high probability. 
But in this case the information about the behavior of the system 
which can be obtained from the readouts of the upper left diagram is 
vague because the band of probable measurement readouts is wide. The 
measurement readout $[E]$ may even remain close to level $E_1$ 
(pointing out the absence of the transition), although the system will 
most probably transit to level $E_2$ just as without any influence of 
the measurement.

The right half of Fig.~\ref{Fig1} refers to the opposite limit. It 
demonstrates the characteristic features of a hard measurement (Zeno 
regime). With high probability the measurement readout $[E]$ shows 
that the system stays on level $E_1$, and the character of the curves 
$[{\cal P}_2]$ confirms this. The hard (Zeno-type) measurement 
influences the measured system very strongly, thus preventing with 
high probability the level transition. The information about the 
development of the system which can be read off from a measured curve 
$E(t)$ of the upper right diagram is much more precise (narrow band of 
probable measurement readouts) and reliable. But the price for this is 
that the Rabi oscillation is completely suppressed.

\subsection{Reliability in the intermediate regime}

For the regime of measurement which is intermediate between Zeno and 
Rabi, the results are also intermediate and much more interesting: The 
influence of the measurement on the system is in this case less than 
in the Zeno regime, and the information given by the measurement 
readout $E(t)$ about the system behavior is more precise than in the 
Rabi regime.

\begin{figure}
\let\picnaturalsize=N
\def\picsize{3.0in}
\def\picfilename{gausfig2.eps}
\ifx\nopictures Y\else{\ifx\epsfloaded Y\else\input epsf \fi
\let\epsfloaded=Y
\centerline{\ifx\picnaturalsize N\epsfxsize \picsize\fi \epsfbox{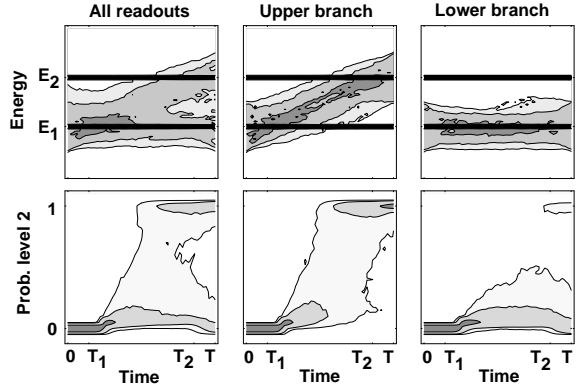}}}\fi
\caption{\label{Fig2}Continuous measurement of a moderate fuzziness
$(4 \pi T_{lr}/T_R=4/3)$. In the left pair of diagrams all measurement
readouts $E(t)$ and curves ${\cal P}_2(t) = |c_2(t)|^2$ are shown.
In the middle and right diagram readouts are selected which point
out that a transition to level 2 happens (middle) or does not
happen (right).
The respective curves ${\cal P}_2(t)$ confirm that the state behaves
like this.}
\end{figure}

The left pair of diagrams in Fig.~\ref{Fig2} shows
all readouts $E(t)$ and the related state curves ${\cal P}_2 (t)$. It 
can be seen that the curves $E(t)$ are separated at final time 
$T$ in two branches approaching the upper or lower energy level. 
The respective probabilities to obtain these curves do not differ 
much. The corresponding effect appears for ${\cal P}_2(t)$.

It is now interesting to compare the curves in more detail in
separating the readouts into two classes. The middle diagrams of
Fig.~\ref{Fig2} refer to readouts pointing to a level transition and the
right diagrams to its prevention. It can be read off directly that in
this regime the information obtained from the readout $E(t)$ is 
rather reliable. This means that a readout $E(t)$ and the
corresponding curve ${\cal P}_2(t)$ for the development of the state
are with high probability in accord with each other. If for
example $E(T)$ is close to level $E_2$, then the most probable
behavior of ${\cal P}_2 (t)$ is characteristic for the transition
from level 1 to level 2. For $E(T)$ close to $E_1$, the system stays
with high probability at level $E_1$ according to ${\cal P}_2 (t)$.
Note that there is an approximately equal probability for a transition
to occur or not to occur.

The numerical analysis has therefore yielded the following important
{\it result}: The visualization of an externally driven quantum transition
by a continuous measurement of energy is possible if the fuzziness
is chosen appropriately, namely to be of intermediate strength.
In this case one may conclude from the actual readout $E(t)$
on the development
in time ${\cal P}_2(t)$ of the state of the system.
But in the intermediate regime it is not possible to predict a single
transition, it may occur or not occur with probabilities of the
same order. This quantum uncertainty is a price for the
visualization of the quantum transition.

\subsection{Dependence on the fuzziness}

\begin{figure}
\let\picnaturalsize=N
\def\picsize{3.0in}
\def\picfilename{gausfig3.eps}
\ifx\nopictures Y\else{\ifx\epsfloaded Y\else\input epsf \fi
\let\epsfloaded=Y
\centerline{\ifx\picnaturalsize N\epsfxsize \picsize\fi \epsfbox{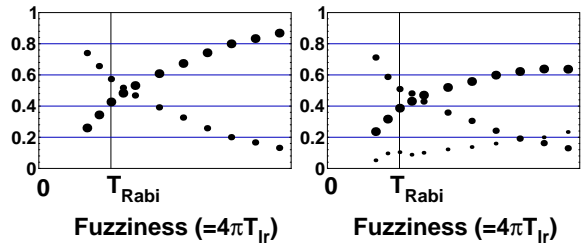}}}\fi
\caption{\label{Fig3}Probabilities that a transition to the upper
level happens (big points) or does not happen (middle sized points)
as functions of the fuzziness. In the left diagram these probabilities are
read off from the curves ${\cal P}_2(t)$, describing the evolution of the
state and in the right diagram from the energy readouts $E(t)$.
Small points present the measurement noise.}
\end{figure}

In Fig.~\ref{Fig3} we present the effects discussed above from a 
different point of view and show how they change continuously for 
increasing fuzziness.  For this purpose we renounce to exhibit the 
total curves $[E]$ and $[{\cal P}_2]$ as functions of time. Instead 
for a given fuzziness represented by the quantity $4 \pi T_{lr}$ we 
divide the curves into two classes, one class 
corresponding to a transition and another to the absence of a 
transition. The probability for a curve to belong to one of the two 
classes is worked out.

In the left diagram of Fig.~\ref{Fig3} we refer to the curves $[{\cal 
P}_2]$ describing the time development of the state of the system. If 
this curve ends close to unity in the sense of ${\cal P}_2(T) > 1/2$, 
we say that the transition has occurred. The big points in the left 
diagram represent the probability for this to happen as a function of 
the fuzziness. For ${\cal P}_2 (T) < 1/2$ the same is indicated by 
middle-size point. The diagram shows how the probability for 
transitions to occur increases with increasing fuzziness. For the 
fuzziness $4 \pi T_{lr}=(10/3)T_R$ at the right end of the diagram the 
probability for a transition is 88 \% (Rabi regime of measurement).

With regard to the energy readout it is assumed in the right part of
Fig.~\ref{Fig3} that the higher level has been reached if $E(T) > E_0 =
(1/2) (E_1+E_2)$. If $E(T) > E_0$ and ${\cal P}_2(T) > 1/2$, then
the readout $[E]$ points to the transition and the real behavior of
the system $[{\cal P}_2]$ is in accord with this information (the
information is valid). The probability of this is presented by big
points at the right part of Fig.~\ref{Fig3}. The middle-sized points
present the probability that $E(T) < E_0$ and ${\cal P}_2(T) < 1/2$ so
that the readout points to the absence of the transition and the real
behavior of the system is in accord with this information (i.e. the
information is valid). In both diagrams the curves and correspondingly
the probabilities are approximately equal for $4\pi T_{lr}/T_R = 4/3$.
This characterizes the intermediate regime. The probability that the
information is false (i.e. the readout points to the transition but
the system stays on the first level or vice versa) is presented by
small points in the right diagram. This type of error is the
measurement noise. This noise increases with fuzziness. For the
intermediate regime the noise is about 20 \%.

\section{Realization of a continuous measurement by a sequence of
fuzzy measurements}\label{realization}

We turn now to concrete experimental setups and consider again a
{\it single quantum system} with two energy
levels $E_1$ and $E_2$ which is under the influence of a driving field.
The total Hamiltonian is $H=H_0+V$. This dynamically developing system
is subject to a continuous fuzzy measurement of energy $H_0$.
In this chapter we want to describe the details of the influence of the
measurement on the system for a rather large class of experimental
realizations. Our intention is to investigate how
their physical microstructure reflects the above given
phenomenological description.

\subsection{The general scheme}\label{realization-general}

The continuous measurement is assumed to be constituted of many 
separated short interactions of duration $\delta \tau$ of the measured 
(two-level) quantum system with another quantum system called the {\it 
meter} to which in this way information about the system is 
transferred. The state of the meter is then measured in a projection 
measurement by a {\it macroscopic measuring apparatus}. The result of 
this measurement is registered. This complete process of interaction 
and projection will be called an {\it elementary observation}. Its 
duration $\delta \tau$ is assumed to be much shorter than the period 
$\tau$ of repetition of such observations $(\delta\tau\ll \tau)$.

The interaction of the measured system with the meter changes its
state depending on the result of the observation. Between two
observations the system evolves according to its Hamiltonian $H$. A
long series of observations leads thus to the evolution of the
two-level system depending on the results of the observations.

To compare this evolution with predictions of the
phenomenological RPI approach as they are presented in
Section~\ref{monitoring} we need the presentation of the measurement
result by a curve $[E]$ called the measurement readout. This curve is
determined, in a specified way, by the results of elementary
observations. This is made in the following way. The whole set of
elementary observations is divided into series, $N$ observations in
each of them with $N$ large enough. One point $E(t)$ of the curve is
elaborated in a specified way (see below) from the results of the 
observations in the $N$-series beginning by the observation at time $t$. 
If the duration $\Delta t=N\tau$ of an $N$-series is short as compared to 
the characteristic time scale of the own dynamics of the system (for
example the Rabi time), then the series of discrete points is a good
approximation for the curve $[E]$. This curve has to be interpreted as
the measurement readout of the continuous measurement in the sense of
the RPI approach. It will be shown that the predictions of the RPI
approach will be in this case valid. 

Fuzziness of the continuous measurement realized in this way
is established in a twofold way:
It depends on the weakness of the individual elementary observations
and on the duration of the intervals between them.
The latter is important too, as the Zeno effect shows,
and must be included as a constitutive element of fuzziness.
We turn now to both aspects in more detail.

\begin{figure}
\let\picnaturalsize=N
\def\picsize{3.0in}
\def\picfilename{gausfig4.eps}
\ifx\nopictures Y\else{\ifx\epsfloaded Y\else\input epsf \fi
\let\epsfloaded=Y
\centerline{\ifx\picnaturalsize N\epsfxsize \picsize\fi \epsfbox{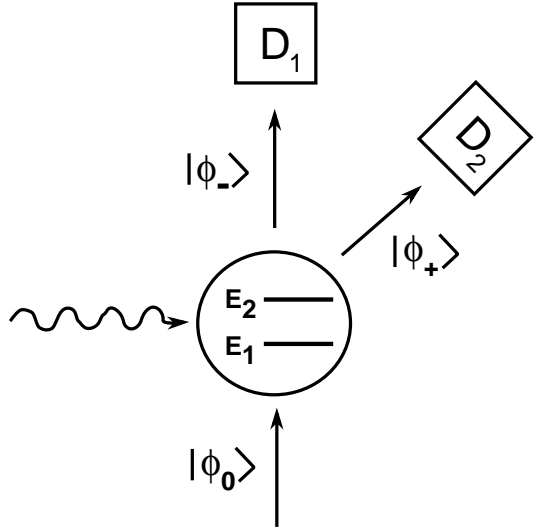}}}\fi
\caption{\label{Fig4}Schematic representation of a possible realization of
an elementary observation.
A two-level system is under the influence of a driving field (wavy line).
Before the observation the states $|\Phi_0\rangle$ of the meter and
$|\psi_0\rangle$ of the system are uncorrelated.
They become entangled during an interaction.
The meter is then measured.
If the state $|\Phi_+\rangle$ is observed (positive result of the
observation), the system is transferred
to the state $|\psi_+\rangle$ which will be close to $|\psi_0\rangle$
if the interaction is weak (for the negative result correspondingly).
The probabilities for the two results depend on the initial
state $|\psi_0\rangle$ of the system.}
\end{figure}

Weakness of the elementary observation means that the system is only 
insignificantly perturbed. The single observation does not result in a 
projection of the state of the system onto an eigenvector of the 
observable (energy in our case). This is realized in the 
following scenario (comp. Fig.~\ref{Fig4}): Before an observation 
which may begin at the time $t$, the system and the meter are 
uncorrelated. The interaction establishes a particular kind of quantum 
correlations between meter and system. Both become entangled in a 
unitary development. The meter is then measured by an apparatus. The 
result is observed. This provides a number as readout and leads to a 
collapse of the state of the combined system. Related to this is  
the transition of the state of the system to a new one which is 
close to the initial state if the coupling strength of the 
interaction is small (weakness of the elementary observation). See 
also Section~\ref{realization-concrete} below. After the elementary 
observation the dynamical development of the system determined by $H$ 
starts again with the new initial state.

Below we will suppose that the meter may transit into one of
only two states as a result of the interaction.
If after the interaction the first state of the meter is observed
in the measuring device, we shall say that the observation
gives a {\it positive result}. If the second state
is observed we call it a negative result. 
Note that we assume throughout the paper that the 
information about the meter state obtained this way is always 
complete (there is no degeneracy). 
This is in analogy to the result of $s$-wave scattering an
electron (taken as meter) by a 2-level system
in \cite{AuMNam97}: it may be deflected (positive result)
or not (negative result) which is indicated by a detector
(measuring apparatus).

To obtain a curve as readout we consider a long sequence
of elementary observations and
divide it into shorter series of $N$ observations
($N$-series). Let the $N$-series beginning at time $t$ contains 
$N_+(t)$ positive and $N-N_+(t)$ negative results. The ratio

\begin{equation}
n(t) = \frac{N_+(t)}{N}
\label{ratio}\end{equation}
will then form a basis for elaborating $E(t)$. The concrete
receipt for this will be given later on (\ref{E-n}).

We turn now to the consideration of the unitary development of
the state of the system interrupted by elementary observations.

\subsection{Change of the state of the system in a series of
elementary observations}\label{realization-change}

We characterize a single elementary observation first. After the 
system has interacted with the meter, subsequently the meter has been 
measured and the result read off, the state of the 2-level system has 
changed. Let the change be of the form

\begin{equation}
\left(
\begin{array}{c}
c_1\\
c_2
\end{array}
 \right) \rightarrow
\left(
\begin{array}{cc}
a_1 & 0 \\
0             & a_2
\end{array}
\right)
\left(
\begin{array}{c}
c_1\\
c_2
\end{array}
 \right)
\label{wave-func}\end{equation}
for the positive observation and the same with $(a^\prime_1, 
a^\prime_2)$ for the negative observation. If the wave function is 
normalized both before and after the interaction, then

\begin{equation}
|a_1|^2 |c_1|^2 + |a_2|^2 |c_2|^2 
= |a'_1|^2 |c_1|^2 + |a'_2|^2 |c_2|^2 = 1.
\label{norm}\end{equation}
It is seen from this formula that the amplitudes $a_i, a^\prime_i$ must
depend on the state $c_i$.

It is more appropriate for our treatment to use wave functions $C_i$
{\em normalized on probabilities} of measurement results. For
this aim let us define amplitudes $u_i$ instead of $a_i$ in such a
way that the transition after the positive observation has the
form

\begin{equation}
\left(
\begin{array}{c}
C_1\\
C_2
\end{array}
 \right) \rightarrow
\left(
\begin{array}{c}
u_1 C_1\\
u_2 C_2
\end{array}
 \right) ;
 \quad
\frac{|u_1|^2 |C_1|^2 + |u_2|^2 |C_2|^2} {|C_1|^2 + |C_2|^2} = p
\label{wave-func-unnorm1}\end{equation}
and the transition after the negative observation is

\begin{equation}
\left(
\begin{array}{c}
C_1\\
C_2
\end{array}
 \right) \rightarrow
\left(
\begin{array}{c}
u'_1 C_1\\
u'_2 C_2
\end{array}
 \right) ;
 \quad
\frac{|u'_1|^2 |C_1|^2 + |u'_2|^2 |C_2|^2} {|C_1|^2 + |C_2|^2} = 1-p.
\label{wave-func-unnorm2}\end{equation}
Here $p$ is the probability for the elementary observation to
give a positive result. 
Weakness of the observation means that its influence on the state
of the system is small. Accordingly all amplitudes
$u_1, u_2, u^\prime_1, u^\prime_2$ will be close to each other.

Now we shall make the important assumption that the amplitudes $u_i, 
u^\prime_i$ do not depend on $c_i$ and therefore on the state of the 
system before the observation. It will be shown later in 
Section~\ref{realization-concrete}, that this assumption is justified 
for a large class of interactions between the two-level system and a 
meter. It is for example valid in the setup described in 
\cite{AuMNam97}. Most important for the analysis of 
elementary observations are moduli of the amplitudes $u_i, u^\prime_i$ 
connected with the parameters $p_1$ and $p_2$

\begin{equation}
p_1 = |u_1|^2 = 1 - |u'_1|^2, \quad p_2 = |u_2|^2 = 1 - |u'_2|^2.
\label{p1p2}\end{equation}
These formulas are obtained from Eq.~(\ref{wave-func-unnorm1}) for $p$ 
for the special cases of $C_2=0$ or 
$C_1=0$ respectively. Therefore $p_1$ (correspondingly $p_2$) is 
the probability that the elementary observation gives the positive 
result if the system is on level 1 (correspondingly 2) before the 
observation. 

Let us now turn to an $N$-series of elementary observations as it was
introduced above.
Eqs.~(\ref{wave-func-unnorm1}),~(\ref{wave-func-unnorm2})
represent the change of the system state due to its interaction with
the meter. In the time between two elementary observations the
evolution of the system is determined by the Hamiltonian $H$
((\ref{2levelEq}) with $\kappa=0$). Because of $\delta \tau \ll
\tau$, the evolution of the system during the period $\tau$ (including
one observation and the time before the next observation, with the
influence of the first observation taken into account) leads to change
of the vector $(C_1, C_2)$ by the following matrix (correspondingly in
the case of the positive or the negative result of the observation):

\begin{equation}
A(\tau) = e^{-iv\tau\sigma_1}
\left(
\begin{array}{cc}
u_1 & 0 \\
0      & u_2
\end{array}
\right)
\quad \mbox{or}\quad
e^{-iv\tau\sigma_1}
\left(
\begin{array}{cc}
u'_1 & 0 \\
0      & u'_2
\end{array}
\right)
\label{HamiltAccount}\end{equation}
where $\sigma_1$ is the Pauli matrix. If we use the basis of 
usual eigenstates of energy $|n\rangle$ instead of the time-dependent 
states (\ref{eigenstates}), the evolution operator $U(\tau)$ will 
differ from $A(\tau)$ in that the exponential in (\ref{HamiltAccount}) 
will be replaced by $e^{-\frac{i}{\hbar}H\tau}$.

For an $N$-series of observations beginning at the time $t$ we will 
treat the $N$ elementary observations as if they had happened directly 
one after the other and take into account the intermediate influence 
of $H$ afterwards adding an exponential in an analogous way.
If in an $N$-series $N_+$ positive results were
obtained in a definite order, then the wave function changes
in such a way that $C_1$ and $C_2$ are multiplied
correspondingly by

\begin{eqnarray}
U_1^{\mathrm{select}}(N_+,N) &=& u_1^{N_+}{u'_1}^{N-N_+}, \nonumber\\
U_2^{\mathrm{select}}(N_+,N) &=& u_2^{N_+}{u'_2}^{N-N_+}
\label{NamplIntermed}\end{eqnarray}
(here we have essentially used the assumption that the amplitudes 
$u_i, u^\prime_i$ do not depend on the state). This is the change of 
the state for the selective description of the measurement when the 
result of each observation in the series is known. But to work out 
$n(t)$ of (\ref{ratio}) it is only necessary to know that there are 
$N_+$ positive results all together. In this case the amplitudes must 
be multiplied by a factor equal to the square root of the number of 
different orders in which positive and negative observations may 
follow (see Appendix)

\begin{eqnarray}
U_1(N_+,N) &=& \sqrt{N\choose{N_+}}u_1^{N_+}{u'_1}^{N-N_+}, \nonumber\\
U_2(N_+,N) &=& \sqrt{N\choose{N_+}}u_2^{N_+}{u'_2}^{N-N_+}.
\label{Nampl}\end{eqnarray}
For the square moduli of these amplitudes we have

\begin{eqnarray}
P_1(N_+,N) = |U_1(N_+,N)|^2
&=& {N\choose{N_+}} p_1^{N_+}(1-p_1)^{N-N_+}, \nonumber\\
P_2(N_+,N) = |U_2(N_+,N)|^2
&=& {N\choose{N_+}} p_2^{N_+}(1-p_2)^{N-N_+}.
\label{NsquarAmpl}\end{eqnarray}
This coincides with the expression for the {\it binomial
distribution}. We will make use of this below.

We can assume, without loss of generality, that the
amplitudes $u_i, u^\prime_i$ have the following phases:

\begin{eqnarray}
u_1 = \sqrt{p_1}\, e^{i\chi},  &&\quad u_2 
= \sqrt{p_2}\, e^{-i\chi},\nonumber\\
u'_1 = \sqrt{1-p_1}\, e^{-i\chi'},  
&&\quad u'_2 = \sqrt{1-p_1}\, e^{i\chi'}.
\label{phases}\end{eqnarray}
Then Eq.~(\ref{Nampl}) may be rewritten as follows:

\begin{eqnarray}
U_1(N_+,N) &=& \sqrt{P_1(N_+,N)} e^{i[N_+\chi-(N-N_+)\chi']},
\nonumber\\
U_2(N_+,N) &=& \sqrt{P_2(N_+,N)} e^{-i[N_+\chi-(N-N_+)\chi']}.
\label{NamplFin}\end{eqnarray}

If the number $N$ of observations taken together in one $N$-series
is large enough, the binomial distribution is close to the
{\it normal distribution} so that expressions (\ref{NsquarAmpl}) may be
with high precision replaced by the {\it Gaussian functions}

\begin{eqnarray}
P_1 &=& {\cal N}_1^2 \exp\left[  
-\frac{N}{2p_1(1-p_1)} (n - p_1)^2 \right],
\nonumber\\
P_2 &=& {\cal N}_2^2 \exp\left[  
-\frac{N}{2p_2(1-p_2)} (n - p_2)^2 \right],
\label{NGauss}\end{eqnarray}
where $n=n(t)$ is defined by Eq.~(\ref{ratio}). Now we have, instead 
of Eq.~(\ref{NamplFin}),

\begin{eqnarray}
U_1 &=&
{\cal N}_1 \exp\left[  -\frac{N(n - p_1)^2}{4p_1(1-p_1)}  
+ iN[n\chi-(1-n)
\chi'] \right],\nonumber\\
U_2 &=&
{\cal N}_2 \exp\left[  -\frac{N(n - p_2)^2}{4p_2(1-p_2)}  
- iN[n\chi-(1-n)
\chi'] \right].\label{NamplGauss}\end{eqnarray}

In Eq.~(\ref{NamplGauss}) we took into account the evolution of 
the two-level system due to its interactions with the meter during 
short elementary observations (each of the duration $\delta\tau$) but not the 
evolution between observations due to the Hamiltonian $H=H_0+V$. The 
latter evolution may be taken into account analogously to 
Eq.~(\ref{HamiltAccount}), by the simple combination of 
Eq.~(\ref{NamplGauss}) and the corresponding exponential. This gives 
in the basis (\ref{eigenstates}) 

\begin{equation}
A(\Delta t) = e^{-iv\Delta t\, \sigma_1}
\left(
\begin{array}{cc}
U_1 & 0 \\
0      & U_2
\end{array}
\right)
\label{NamplGaussEvolA}\end{equation}
and in the conventional basis $|n\rangle$ 

\begin{equation}
U(\Delta t) = e^{-\frac{i}{\hbar}H\Delta t}
\left(
\begin{array}{cc}
U_1 & 0 \\
0      & U_2
\end{array}
\right).
\label{NamplGaussEvol}\end{equation}
These formulas are obtained in the approximation when the 
commutator of $H$ with the evolution operator 
(\ref{wave-func-unnorm1}), (\ref{wave-func-unnorm2}) is neglected. 
This commutator is small because the differences $u_1-u_2$ and 
$u^{\prime}_1-u^{\prime}_2$ are small in the case of a fuzzy 
observation. A more detailed consideration shows that the commutator 
may be neglected if the $N$-series is not too long so that 

\begin{equation}
N^2 \ll \frac{\sqrt{p_0}}{\Delta p v\tau}.
\label{Nrestrict}\end{equation}
where $p_0$ and $\Delta p$ denote correspondingly (\ref{p0Dp}) the 
arithmetic mean of and the difference between the two probabilities 
$p_1$ and $p_2$. The parameter $v\tau=\pi\tau/T_R$ is assumed 
to be small. Later we shall see 
that the ratio $\Delta p^2/p_0$ is small if $\tau$ is less than the 
level resolution time $T_{\mathrm lr}$ (see (\ref{cont-approx})). 
Therefore, the number $N$ may be large even despite of the 
restriction~(\ref{Nrestrict}).

Eq.~(\ref{NamplGaussEvol}) has the following physical 
meaning: If for a $N$-series of duration $\Delta t$ the number $n(t)$ 
of (\ref{ratio}) is registered, then the state of the system will have 
been evolved according to $|\psi (t+\Delta t) \rangle = U(\Delta t) 
\cdot |\psi (t) \rangle$. This evolution is not unitary. The 
primary reason for this is the non-unitary character of the operators 
(\ref{wave-func-unnorm1}), (\ref{wave-func-unnorm2}) describing 
transitions due to elementary observations.

\subsection{Comparison with the phenomenological approach}
\label{realization-comparison}

Our general experimental setup leads to $n(t)$ of (\ref{ratio}) as a
readout.  The remaining problem is therefore, how can we obtain from
$n(t)$ a function $E(t)$ which represents the energy readout in the
phenomenological scheme of Section~\ref{SectPhenomen}. In answering this
question we will at the same time obtain a new physical
interpretation of the phenomenological $E(t)$.

For this aim let us discuss the statistics of the results in a
single $N$-series of observations, starting from the state $|\psi
(t)\rangle$ characterized by $(C_1, C_2)$.
The probability for the number of positive results in the
$N$-series to be $N_+$ is (see Appendix)

\begin{equation}
P(N_+,N) = {N\choose{N_+}} \left[
{\cal P}_1(t) \, p_1^{N_+}(1-p_1)^{N-N_+}
+{\cal P}_2(t) \, p_2^{N_+}(1-p_2)^{N-N_+}
\right]
\label{ProbN+}\end{equation}
where

\begin{equation}
{\cal P}_i = \frac{|C_i|^2}{|C_1|^2 + |C_2|^2}
\label{calP12}\end{equation}
is the probability to be on levels $i$ when the $N$-series start.
The mean value of $n$ is therefore

\begin{equation}
\overline{n}=\sum_{N_+=0}^N \frac{N_+}{N}P \left(N_+,N \right)
=  p_1{\cal P}_1(t) + p_2{\cal P}_2(t).
\label{mean-n1}\end{equation}

According to (\ref{wave-func-unnorm1}), this is equal to
the probability $p$ to obtain a positive observation
in the state $|\psi(t)\rangle$.
This leads to

\begin{equation}
\overline{n}=p.
\label{mean-n2}\end{equation}

We have thus shown that the probability to obtain a positive
result in a single elementary
observation of a given state $|\psi (t)\rangle$ of the system
agrees with the mean value $\overline n$ of
$n=N_+/N$ with $n$ being attributed to a complete $N$-series
starting with this state.
In this sense
the actually measured $n(t)$ may be regarded as the best possible
estimate, on the basis of $N$ elementary observations, of the
mean value $\overline n(t)$ and therefore of $p(t)$. The
quantum-mechanical mean value $\overline{E}$ of the energy in
the state $|\psi(t)\rangle$ is

\begin{equation}
\overline{E} = E_1\, {\cal P}_1 + E_2\, {\cal P}_2.
\label{meanE}\end{equation}
Comparison with (\ref{mean-n1}) shows that the mean values
$\overline{n}$ and $\overline{E}$ are related according to

\begin{equation}
\frac{\overline{E} - E_0}{ {\Delta E}} =
\frac{\overline{n}-p_0}{{\Delta p}}
\label{Ebar-nbar}\end{equation}
where the following notations have been used:

\begin{equation}
\quad p_0 = \frac 12 (p_1 + p_2), \quad
{\Delta p} = p_2 - p_1
\label{p0Dp}\end{equation}
($E_0$ and ${\Delta E}$ have already been introduced by analogous
formulas).

Eq.~(\ref{Ebar-nbar}) gives a hint how in a continuous measurement
an energy readout $E(t)$ may be related with the
function $n(t)$ obtained from the observations. Transcribing
(\ref{Ebar-nbar}) we try in the following

\begin{equation}
\frac{E(t) - E_0}{{\Delta E}} = \frac{n(t)-p_0}{{\Delta p}}
\label{E-n}\end{equation}
as definition of $E(t)$. Herewith Eq.~(\ref{NamplGauss}) may be
rewritten as (we omit the argument $t$):

\begin{eqnarray}
U_1 &=&
{{\cal N}_1} \exp\left[
 -\frac{N{\Delta p}^2}{4 p_1(1-p_1)}
\left( \frac{E - E_1}{{\Delta E}} \right)^2 \right]
\exp\left[ iN \left( -\chi'
+ (\chi+\chi')\left(p_0+{\Delta p}\frac{E-E_0}{{\Delta E}} 
\right) \right) \right] ,
\nonumber\\
U_2 &=&
{{\cal N}_2} \exp\left[
 -\frac{N{\Delta p}^2}{4 p_2(1-p_2)}
\left( \frac{E - E_2}{{\Delta E}} \right)^2 \right]
\exp\left[ -iN \left( -\chi'
+ (\chi+\chi')\left(p_0+{\Delta p}\frac{E-E_0}{{\Delta E}} 
\right) \right) \right] .
\label{NamplGaussE}\end{eqnarray}

Let us consider in more detail the case when the second exponential
(the phase factor) in each of the expressions (\ref{NamplGaussE}) may
be neglected (this assumptions are valid for the system considered in 
\cite{AuMNam97}). Besides, assume that $p_1$ and $p_2$ are close enough 
so that $p_1(1-p_1)$ and $p_2(1-p_2)$ in Eq.~(\ref{NamplGaussE}) may be 
taken to be equal to each other and to $p_0(1-p_0)$ (this may be shown is 
the case if $N(p_1+p_2)\gg 1$). Then

\begin{equation}
U_i =
{\cal N}_0 \exp\left[
 -\kappa \Delta t
( E - E_i)^2
\right]
=
{\cal N}_0 \exp\left[
 -\frac{\Delta t}{T_{\mathrm lr}}\
\left( \frac{E - E_i}{{\Delta E}} \right)^2
\right]
\label{NamplGaussReal}\end{equation}
where $\Delta t=N \cdot \tau$ is the duration of
the $N$-series and we have introduced the abbreviation

\begin{equation}
T_{\mathrm lr} = \frac{1}{\kappa{\Delta E}^2} 
= \tau\frac{4p_0(1-p_0)}{{\Delta p}^2}.
\label{Tlr-pi}\end{equation}

Substituting the expressions (\ref{NamplGaussReal}) for $U_i$ in 
Eq.~(\ref{NamplGaussEvol}) and normalizing properly, we obtain 
finally for the evolution up to the end of the $N$-series

\begin{equation}
U(\Delta t)
= \exp\left[\left(-\frac{i}{\hbar}H -\kappa (E-H_0)^2\right)
\Delta t\right].
\label{GaussEvolFin}\end{equation}
The evolution during a long period [0,T], which is
divided into many
$N$-series, is described by the product of the respective amplitudes
of the form (\ref{GaussEvolFin}). Because of $\Delta t \ll T$ the sum
in the exponent may then be replaced by an integral and the discrete
values $E(t)$ join together to a curve $[E]$.  Therewith we have
found the dynamics which agree completely with the one of the
effective Hamiltonian (\ref{effect-Ham}).

Accordingly we have obtained the {\it result} that the scheme 
discussed above realizes exactly the continuous fuzzy measurement of 
energy as it is described phenomenologically by the RPI approach or a 
complex Hamiltonian, provided the readout curve $[E]$ is obtained on 
the basis of the values $E(t)$ worked out according to (\ref{E-n}). In 
this sense the $E(t)$ may now be regarded as the energy readout. At 
the same time $T_{lr}$ of (\ref{Tlr-pi}) is nothing but the previously 
introduced level resolution time.

It is seen from Eq.~(\ref{Tlr-pi}) that a large level resolution 
time $T_{\mathrm lr}$ (and therefore a fuzzy continuous measurement) 
is provided by small $\Delta p$ and large $\tau$. This means that each 
elementary observation must be fuzzy and the repetition of 
observations not too often. The series of elementary observations may 
be considered as a continuous measurement if the period $\tau$ between 
single observations is much less than the characteristic time scale of 
the continuous measurement $T_{\mathrm lr}$. We see from 
Eq.~(\ref{Tlr-pi}) that this is provided if 

\begin{equation} 
\frac{p_0}{\Delta p^2}\gg 1
\label{cont-approx}\end{equation}
(we assumed that $1-p_0$ is of the order of unity). This makes it 
possible to choose large number $N$ in a $N$-series despite of the 
restriction (\ref{Nrestrict}).

Having established in (\ref{E-n}) how the energy readout $[E]$
appearing in the phenomenological scheme can be obtained in practice,
we may now read off its interpretation:  The construction above shows
that a point $E(t)$ of the readout $[E]$ in the
phenomenological approach may be regarded as the
estimate of the usual quantum-mechanical mean value
$\overline E$ of the energy of
the system at time $t$.
This is a remarkable result because $\overline E$ refers as mean value
to a large number of sharp energy measurements (projection measurements)
on an ensemble that is characterized by the state $|\psi (t)\rangle$ the
system assumes at time $t$. Operationally this
has nothing to do with a continuous measurement (or a
sequence of measurements) where the state of the system
changes in time under the measurement.
The indicated correspondence between $E(t)$ and $\overline E$
means roughly speaking the following:
Because of the weakness of an elementary observation it does not change
the state significantly so that a succession
of observations leads to similar results as many sharp measurements
performed on the same state.
But note that $E(t)$ is only an estimate of $\overline E$ 
obtained from a finite number of observations. In general they
will differ.
If the precision of this estimation is poor corresponding to the
weakness of the measurement, then the actual value $E(t)$ obtained
does not have to be close
to $\overline E$. This is a reflection of fuzziness.

\subsection{A concrete model}\label{realization-concrete}

Our aim is now to discuss in more detail the assumptions made above in
Sections~\ref{realization-change},~\ref{realization-comparison} and to
show that they are reasonable by sketching that they are fulfilled
for a rather large class of interactions between a 2-level system
and the meter.

Let our system with Hamiltonian $H=H_0+V$ interact with the meter
with Hamiltonian $H_M$ via the interaction $V_I$. Then the
Schr\"odinger equation for the {\it composed quantum system} (the 2-level
system plus the meter) with state $|\Psi\rangle$ in the product space is

\begin{equation}
i\hbar|\dot \Psi\rangle = (H+H_M+V_I) |\Psi\rangle.
\label{Schroed-eq}\end{equation}

We turn to a single elementary observation (comp. Fig.~\ref{Fig4}).
System and meter are
initially uncorrelated so that the state $|\Psi(0)\rangle$
of the combined system is a product of the system state
$|\psi_0 \rangle$ and the normalized meter state $|\Phi_0 \rangle$

\begin{equation}
|\Psi(0)\rangle = |\psi_0\rangle |\Phi_0\rangle.
\label{State-prod}\end{equation}
Restricting to a time independent interaction operator $V_I$ and assuming
that the duration $\delta \tau$ of the interaction is short so that
the dynamics due to $H+H_M$ is negligible during this period, we
have for the state at the end of the interaction as consequence of the
unitary development

\begin{equation}
|\Psi({\delta\tau})\rangle = \left( 1 - \frac{i}{\hbar}  
V_I{\delta\tau} \right) |\Psi(0)\rangle.
\label{Psi-dt}\end{equation}

The states of the meter corresponding to a positive or negative
result of the observation are denoted by $|\Phi_+\rangle$ and
$|\Phi_- \rangle$ respectively. They form a basis in the meter
space. Then, multiplying Eq.~(\ref{Psi-dt}) by

\begin{equation}
1=|\Phi_+\rangle \langle\Phi_+| + |\Phi_-\rangle \langle\Phi_-|,
\label{basis}\end{equation}
we have

\begin{equation}
|\Psi({\delta\tau})\rangle = |\Phi_+\rangle |\psi_+\rangle +  |\Phi_-\rangle |\psi_-\rangle
\label{Psi-dt-expand}\end{equation}
where

\begin{eqnarray}
|\psi_+\rangle &=& \langle\Phi_+| 
\left( 1 - \frac{i}{\hbar} {\delta\tau} V_I \right)
|\Phi_0\rangle|\psi_0\rangle  , \nonumber\\
|\psi_-\rangle &=& \langle\Phi_-| 
\left( 1 - \frac{i}{\hbar} {\delta\tau} V_I \right)
|\Phi_0\rangle |\psi_0\rangle . 
\label{psi-plus-min}\end{eqnarray}
The interaction converts the composed system into the entangled state 
(\ref{Psi-dt-expand}). Measuring the subsystem meter by a macroscopic 
measuring apparatus transfers the 2-level system to one of the states 
of (\ref{psi-plus-min}), depending on whether the positive or negative 
result is registered.

We introduce now that the interaction $V_I$ is not only of short 
duration $\delta \tau$ but also weak. We assume for simplicity that in 
case of a negative result of the observation the final state of the 
meter coincides with its initial state, 
$|\Phi_-\rangle=|\Phi_0\rangle$. This is fulfilled for the example 
given in \cite{AuMNam97} based on scattering, where the outgoing state 
of an electron for the negative result of the observation (no 
deflection) agrees with its ingoing state (before scattering). 
The consequence is the orthogonality of $|\Phi_0\rangle$ and $|\Phi_+ 
\rangle$ so that we have up to a phase factor $(-i)$

\begin{eqnarray}
|\psi_+\rangle &=& \frac{{\delta\tau}}{\hbar} 
\langle\Phi_+| V_I  |\Phi_0\rangle|\psi_0\rangle  ,
\nonumber\\
|\psi_-\rangle &=& \left( 1 - \frac{i}{\hbar} 
{\delta\tau}  \langle\Phi_0|V_I |\Phi_0\rangle
\right)|\psi_0\rangle .
\label{psi-plus-min-simple}\end{eqnarray}

To represent fuzziness we have made above the assumption that $p_1$ 
and $p_2$ are close. This can be realized by choosing an interaction 
operator $V_{I}$ which has the same structure but slightly 
different coupling constants for the different levels of the measured 
system:

\begin{equation}
V_{I} = V_{I0} (1- g\, \sigma_3).
\label{InteractHam}\end{equation}
$\sigma_3$ is here the Pauli matrix acting in the two-dimensional 
space of levels 1 and 2 and $g$ is a number much smaller than 1. 
Then the final states of the measured system after the short 
interaction with the meter are

\begin{eqnarray}
|\psi_+\rangle &=& a (1- g\, \sigma_3) |\psi_0\rangle  ,
\nonumber\\
|\psi_-\rangle &=& e^{ibg\sigma_3}
 |\psi_0\rangle
\label{psi-plus-min-sigma}\end{eqnarray}
where

\begin{equation}
a =\frac{{\delta\tau}}{\hbar} \langle\Phi_+| V_{I0}  |\Phi_0\rangle , 
\quad
b = \frac{{\delta\tau}}{\hbar}  \langle\Phi_0|V_{I0} |\Phi_0\rangle.
\label{ab}\end{equation}
Comparing Eq.~(\ref{psi-plus-min-sigma}) with 
(\ref{wave-func-unnorm1}) and (\ref{wave-func-unnorm2}) we can infer 
from (\ref{p0Dp})

\begin{equation}
p_0 = |a|^2(1 + g^2), \quad {\Delta p} = 4g|a|^2.
\label{parameters}\end{equation}
Probabilities $p_1$ and $p_2$ are therefore close, 
if $g$ is chosen to be small enough.

The level resolution time of (\ref{Tlr-pi}) becomes

\begin{equation}
T_{\mathrm lr} = \frac{\tau}{4g^2|a|^2}.
\label{Tlr-concrete}\end{equation}
Thus, $1/4g^2|a|^2$ is a critical number of observations, leading to the
resolution of energy levels (see a detailed discussion of this
resolution in \cite{AudM97}). The phase factor in
Eq.~(\ref{psi-plus-min-sigma}) is negligible for the continuous
measurement of the duration $T$ if 

\begin{equation} 
g\,b \ll \frac{\tau}{T}.
\label{phase-neglig}\end{equation}
This is valid for $T$ of the order of $T_{\mathrm lr}$ if  $b\ll 4g|a|^2$. 
This condition is fulfilled for the system considered in 
\cite{AuMNam97}. 

Thus the quantum-mechanical analysis of a simple model of a
short observation of a two-level system proves that the resulting
evolution of the system is in accord with the scheme postulated in
Section~\ref{realization-change}. Together with the consideration of
the whole Section~\ref{realization} this proves that the continuous
quantum measurement described phenomenologically in
Section~\ref{SectPhenomen} can be realized in practice. The meter system
appropriate for such a realization may be chosen from a rather wide
class considered above. Moreover, this class may even be widened. In
fact the only essential condition for the realization is that
the interaction with the meter be weak (and observations not too
often).

\section{Discussion}\label{sectDiscuss}

We have obtained the following results:

\begin{enumerate}
\item It is possible to visualize the level transition of an
externally driven single two-level system by a continuous measurement
of energy, if its fuzziness is appropriately chosen. It is most
adequate to show this within the phenomenological RPI-approach or to
use the equivalent Schr\"odinger equation with complex Hamiltonian.
The information obtained about the level transition has then a 
restricted but not too small reliability (measurement noise is about
20\%) and may be obtained only at the price of changing the
probability of the transition. 
\item Measurements of this type may be realized by a long
series of sufficiently weak interactions of the system with
a quantum mechanical meter system, which is then measured
by a macroscopic apparatus.
Fuzziness may be established this way.
The discussion of the microphysical dynamics shows that the model 
reproduces the phenomenologically obtained results. The class of 
experimental setups which enable this realization is rather large.
\item The microphysical investigations lead to operational 
interpretations of phenomenological quantities as for example the 
energy readout $E(t)$. It turns out to be the
estimate of the mean value
$\overline E(t)$ of the energy of the system with respect to its state
$|\psi(t)\rangle$ at that time.
That both energies must not agree exactly, leaves room for fuzziness.
\item Besides weakness of interactions with the meter,
visualization of a quantum transition requires to neglect details of
information. This is reflected 1)~in the scheme for working out the
only one number $E(t)$ from a long series of interactions with the
meter and 2)~in the subsequent smoothing of the function ${E}$. It is
only after these two stages of roughening of the measurement results
that they can be interpreted as the visualization of the transition.
\end{enumerate}

As far as an experimental realization is concerned, we add: 
The realization of the continuous measurement by a long series of 
elementary observations as described in Sect.~\ref{realization} 
generalizes the experimental setups 
proposed in \cite{BruHar90,AuMNam97}. However the aim of the 
paper \cite{BruHar90} (see also \cite{Schenzle92} for more detailed 
discussion of the same setup) was not the investigation of the 
continuous measurement in general, but the decoherence of levels 
resulting if the measurement is long enough (in our notations, longer 
than decoherence time, $T\gg T_{\rm lr}$). Note that we referred to 
a different domain when discussing in Sect.~\ref{monitoring} how a driven 
level transition may be monitored by an appropriate continuous 
measurement. 

We close with two remarks: The scheme of the realization, though
general enough, may be further generalized.
In fact, almost any weak and short interaction with
the two-level system which does not disturb its state significantly
may be used to realize the fuzzy continuous measurement of energy.

It is interesting that for a wide class of elementary observations their
repetition leads to the simplest form of the continuous measurement
characterized by a Gaussian functional in the restricted path
integral or by a quadratic imaginary term in the
Hamiltonian. The resulting measurement is characterized
by only one single parameter, namely fuzziness.  
The interaction of the measured system with
the meter may depend on many parameters, but only one combination of
them, fuzziness, is important for the description of a resulting
continuous measurement. This is quite analogous to the the well-known 
fact from probability theory (proved in the central limiting theorem) 
that the result of a long series of random events is normally distributed 
independently of the distributions of single events in the series. Therefore, 
consideration in Sect.~\ref{realization} formulates and proves the simplest 
quantum analog of the classical central limiting theorem. 

\acknowledgements
The authors acknowledge helpful discussions with F. Burgbacher, Th. Konrad 
and V. Namiot. This work was supported in part by the Deutsche 
Forschungsgemeinschaft and by the Russian Foundation for Basic Research, 
grant 98-01-00161.

\appendix
\section*{}
\label{SectStatistics}

To prove Eqs.~(\ref{Nampl}) and (\ref{ProbN+}) we start with
(\ref{NamplIntermed}) which describes the development of a state
under the influence of $N$ observations if the succession (the order)
of the negative or positive results in the $N$-series is known.  We
write this in the form

\begin{equation}
|\psi\rangle \rightarrow
U_{\mathrm{select}}({\mathrm{order}}) |\psi\rangle 
= U_{\mathrm{select}}
|\psi\rangle
\label{select-ampl}\end{equation}
where $U_{\mathrm select}$ (order) is a
selective amplitude of (\ref{NamplIntermed}).

If the order of the positive and negative results is not known or
not taken into account we have to sum up over all alternatives
in terms of a density matrix,
\begin{eqnarray}
|\psi\rangle \langle\psi |  &\rightarrow &
\sum_{{\mathrm{order}}}
U_{\mathrm{select}}({\mathrm{order}}) |\psi\rangle \langle\psi |
U_{\mathrm{select}}^{\dagger}({\mathrm{order}}) \nonumber\\
&=& {N\choose{N_+}} U_{\mathrm{select}} |\psi\rangle \langle\psi |
U_{\mathrm{select}}^{\dagger}.
\label{nonselect-dens}\end{eqnarray}
The summation here has reduced to a multiplication by the
number of different orders because the amplitude
(\ref{NamplIntermed}) does in fact not depend on the orders. Going
back to the evolution of the state we obtain

\begin{equation}
|\psi\rangle \rightarrow \sqrt{{N\choose{N_+}}}U_{\mathrm{select}} 
|\psi\rangle,
\label{nonselect-ampl}\end{equation}
and therefore Eq.~(\ref{Nampl}) is proved.

The probability $P(N_+, N)$ to have $N_+$ positive results in $N$
observation (independently of the order of these results) may be found
as a trace of the r.h.s. of Eq.~(\ref{nonselect-dens}) if the vector
$|\psi\rangle$ is normalized.
If it is not normalized, the resulting expression must be divided by the
square of its norm $||\psi||^2=|C_1|^2+|C_2|^2$.
With the expressions (\ref{NamplIntermed}) for the amplitude $U_{\mathrm 
select}$ and making use of (\ref{NsquarAmpl}) we obtain the
probability $P(N_+, N)$ in the form (\ref{ProbN+}).

\end{document}